\documentclass[12pt]{article}

\usepackage{graphicx}
\begin{document}

\begin{center}

{\bf The shadow of M87* black hole within rational nonlinear electrodynamics} \\
\vspace{3mm}
S. I. Kruglov
\footnote{E-mail: serguei.krouglov@utoronto.ca}

\vspace{3mm}
\textit{Department of Physics, University of Toronto, \\60 St. Georges St.,
Toronto, ON M5S 1A7, Canada\\
Department of Chemical and Physical Sciences, University of Toronto,\\
3359 Mississauga Road North, Mississauga, ON L5L 1C6, Canada} \\
\vspace{5mm}

\end{center}
\begin{abstract}
We consider rational nonlinear electrodynamics with the Lagrangian ${\cal L} = -{\cal F}/(1+2\beta{\cal F})$
(${\cal F}=(1/4)F_{\mu\nu}F^{\mu\nu}$ is the Lorentz-invariant), proposed in \cite{Krug2}, coupled to General Relativity. The effective geometry induced by nonlinear electrodynamics corrections are found. We determine shadow's size of regular non-rotating magnetic black holes and compare them with the shadow size of the super-massive M87* black hole imaged by the Event Horizon Telescope collaboration. Assuming that the black hole mass has a pure electromagnetic nature, we obtain the black hole magnetic charge. The size of the shadow obtained is very close to the shadow size of non-regular neutral Schwarzschild black holes. As a result, we can interpret the super-massive M87* black hole as a regular (without singularities) magnetized black hole.
\end{abstract}

\section{Introduction}

The black hole (BH) shadows are formed due to gravitational lensing near a BH event horizon because the gravitational field is very strong and photon orbits are unstable forming orbits which make the photon sphere which is not necessarily circular. The shadow separates capture orbits and scattering orbits of photons around BHs \cite{Luminet}-\cite{Narayan}.
The size and shape of the BH shadow strongly depends on the BH mass and distance but very weakly on its spin. For a non-rotating black hole the shape of a shadow is a circle, whereas for a BH with a spin the shadow will be deformed due to dragging effects.
Very long baseline interferometry can detect the shadows of super-massive black holes (SMBHs)  \cite{Falcke}, \cite{Event1}. Event Horizon Telescope (EHT) collaboration detected the shadow of M87$^*$ at the center of the elliptical galaxy Messier 87, \cite{Event1}. The image of the M87$^*$ shadow is consistent with the Kerr BH in General Relativity (GR). At the same time BH shadows can test deviations from GR \cite{Israel}-\cite{Liu}. It is possible to extract an information from the M87$^*$ shadow, regarding properties of the BH such as constraints on fundamental physics, see e.g. \cite{Wei}-\cite{Vagnozzi2}. For a Schwarzschild (non-rotating) BH, the shadow diameter is $3\sqrt{3}r_{Sch}$, where $r_{Sch}=2MG$ is the Schwarzschild radius \cite{Synge}.
Gravitational collapse in GR leads to the undesirable existence of singularities \cite{Penrose}, \cite{Hawking}.
The cosmic censorship conjecture states that all singularities of gravitational collapse are hidden by event horizons of BHs, and should not be naked \cite{Penrose1}. But it is desirable to have solutions which
avoid singularities.

Some nonlinear electrodynamics (NLED) represent models having regular BH solutions. The
first example of NLED is Born$-$Infeld electrodynamics \cite{Born} possessing a singularity at $r=0$, despite a finite self-energy of an electric charge. Another example of non-regular NLED is Euler$-$Heisenberg electrodynamics with the action of quantum electrodynamics (QED) taking into account loop corrections \cite{Heisenberg}. NLED models coupled to GR and their BH solutions were studied in (an incomplete list) \cite{Pellicer}-\cite{Kruglov11}. It was shown in \cite{Bronnikov} that NLED coupled to GR does not possess a static, spherically symmetric solution with a regular center and non-zero electrical charge if NLED has the Maxwell asymptotic at weak-field limit. Therefore, only magnetically charged BH can have regular solution when NLED becomes Maxwell electrodynamics at weak fields.

In  this paper we consider non-spinning BHs and investigate regular solutions carrying a magnetic charge within rational NLED \cite{Krug2} coupled to GR. By computing the shadow size within our model and compare it with the size of the shadow of the M87$^*$ BH, we evaluate the magnetic charge of the BH. It worth noting that the shadow size is independent of the spin of the BH with an uncertainty $\leq10\%$ \cite{Event1}.

\section{The model}

Let us consider the motions of photons in a regular spacetime which being a solution to the Einstein theory coupled to rational NLED. This will allow us to determine the shadow size of magnetically charged BH. The light propagates in NLED along the null geodesics and induces the effective geometry modifying original background spacetime \cite{Novello}, \cite{Novello1}.

We consider the rational NLED, proposed in \cite{Krug2}, which is converted to linear Maxwell's electrodynamics  at the weak field limit ($\beta {\cal F} \to 0$) and it is described by the Lagrangian
\begin{equation}
{\cal L} = -\frac{{\cal F}}{1+2\beta{\cal F}}.
 \label{1}
\end{equation}
The parameter $\beta$ is positive and possesses the dimension of (length)$^4$, ${\cal F}=(1/4)F_{\mu\nu}F^{\mu\nu}=(B^2-E^2)/2$, $F_{\mu\nu}$ is the field tensor. The action of GR coupling to NLED is given by
\begin{equation}
S=\int d^4x\sqrt{-g}\left(\frac{1}{16\pi G}R+ {\cal L}\right),
\label{2}
\end{equation}
where $G$ is the Newton constant. The variation of the action (2) with respect to the metric tensor gives the Einstein equation
\begin{equation}
R_{\mu\nu}-\frac{1}{2}g_{\mu\nu}R=-8\pi GT_{\mu\nu}=8\pi G\left(\frac{F_\mu^{~\alpha}F_{\nu\alpha}}{(1+2\beta{\cal F})^{2}}+g_{\mu\nu}{\cal L}\right).
\label{3}
\end{equation}
In order to solve the field equations, we use the ansatz of the static and spherically symmetric metric with the squared of the line element
\begin{equation}
ds^2=-f(r)dt^2+\frac{1}{f(r)}dr^2+r^2(d\vartheta^2+\sin^2\vartheta d\phi^2).
\label{4}
\end{equation}
The metric function of the magnetic BH within rational NLED is given by \cite{Kr}
\[
f(x)=1-\frac{q_m^{3/2}G}{4\sqrt{2\beta}x}\biggl(\ln\frac{x^2-\sqrt{2}x+1}{x^2+\sqrt{2}x+1}
\]
\begin{equation}
+2\arctan(\sqrt{2}x+1)-2\arctan(1-\sqrt{2}x)\biggr),
\label{5}
\end{equation}
where we introduced the unitless variable $x=r/\sqrt[4]{\beta q_m^2}$ ($q_m$ is the magnetic charge which is unitless in our Gaussian units with $c=1$). It worth noting that the magnetic charge $q_m$ is an integration constant of the Einstein equation. As $r\rightarrow \infty$ the metric function (5) reduces to the Reissner$-$Nordstrom solution with corrections depending on $\beta$ \cite{Kr}. If the Schwarzschild mass is nonzero the curvature invariants possess singularities. To have the regular BH we ignore the Schwarzschild mass so that the total mass of the BH is the magnetic mass \cite{Kr}
\begin{equation}
m_M=\frac{\pi q_m^{3/2}}{4\sqrt{2}\beta^{1/4}}\approx 0.56\frac{q_m^{3/2}}{\beta^{1/4}}.
\label{6}
\end{equation}
Equation (5) can be rewritten as
\[
f(x)=1-\frac{Bg(x)}{x},~~~B=\frac{q_mG}{\sqrt{\beta}},
\]
\begin{equation}
g(x)=\frac{1}{4\sqrt{2}}\left(\ln\frac{x^2-\sqrt{2}x+1}{x^2+\sqrt{2}x+1}
+2\arctan(\sqrt{2}x+1)-2\arctan(1-\sqrt{2}x)\right),
\label{7}
\end{equation}
where the $B$ is the unitless constant (do not confuse the constant $B$ with the magnetic field). We will explore the method of \cite{Novello}, \cite{Novello1} to describe the effective geometry induced by our NLED. Photons propagate along the null geodesics of this effective geometry.
The null geodesics of the photon paths are described by the effective spacetime
with the static and spherically symmetric metric \cite{Novello}, \cite{Novello1}, \cite{Vagnozzi}
\begin{equation}
g^{\mu\nu}_{eff}={\cal L}_{{\cal F}}g^{\mu\nu}-{\cal L}_{{\cal F}{\cal F}}F^\mu_\alpha F^{\alpha\nu},
\label{8}
\end{equation}
where ${\cal L}_{{\cal F}}=\partial {\cal L}/\partial {\cal F}$.
The effective metric for photons found from Eqs. (4), (7) and (8) is given by
\begin{equation}
ds_{eff}^2=A(r)\left(-f(r)dt^2+\frac{1}{f(r)}dr^2\right)+h(r)r^2(d\vartheta^2+\sin^2\vartheta d\phi^2),
\label{9}
\end{equation}
where
\begin{equation}
A(x)=-{\cal L}_{\cal F}=\frac{x^8}{(x^4+1)^2} ,~~~h(x)=-{\cal L}_{\cal F}-{\cal L}_{{\cal F}{\cal F}}\frac{q_m^2}{r^4}=\frac{x^8(x^4-3)}{(x^4+1)^3}.
\label{10}
\end{equation}
The metric functions $A(r)$ and $h(r)$ ($r=x\sqrt[4]{\beta q_m^2}$) must be positive in order that the effective geometry does not change its signature during the photon motion. This leads to the requirement $x>\sqrt[4]{3}\approx 1.32$ or $r> r_{eff}\equiv \sqrt[4]{3}\beta^{1/4}\sqrt{q_m}$. Another restriction is $r>r_h$, where $r_h$ is the radius of the event horizon ($f(r_h)=0$). When $r_{eff}< r_h$ the exterior region of the BH is realized by $r > r_h$ but if $r_{eff}> r_h$ the range of the photon motion outside of the BH is given by $r > r_{eff}$. The equation of motion along the equatorial plane with $\vartheta=\pi/2$ for a null geodesic in terms of an effective potential $V(r)$, reads \cite{Vagnozzi}
\begin{equation}
\left(\frac{dr}{d\phi}\right)^2=V(r)=r^4\left(\frac{E^2h(r)^2}{L^2A(r)^2}-\frac{f(r)h(r)}{A(r)r^2}\right),
\label{11}
\end{equation}
where $E=f(r)A(r)\dot{t}$ and $L=r^2h(r)\dot{\phi}$ are the photon constants of motion, the photon total
energy and angular momentum, respectively. Unstable circular orbits take place when $dV/dr=V=0$. Another condition $d^2V/dr^2>0$ should be satisfied for stable orbits. Then Eq. (11) with $dV/dr=V=0$ leads to
\[
b^{-2}\equiv\frac{E^2}{L^2}=\frac{f(r)A(r)}{r^2h(r)},
\]
\begin{equation}
2f(r)A(r)h(r)+rf(r)h'(r)A(r)-rf'(r)A(r)h(r)-rf(r)h(r)A'(r)=0,
\label{12}
\end{equation}
where $b$ is the impact parameter and the prime means the derivative with respect to $r$.
Making use of Eqs. (5), (10) and (12) we obtain relations as follows:
\[
b^{-2}=\frac{1}{q_m\sqrt{\beta}}\left(1-\frac{Bg(x)}{x}\right)\frac{x^4+1}{x^2(x^4-3)},
\]
\begin{equation}
2x(x^8+6x^4-3)-Bg(x)(3x^8+10x^4-9)+Bx^3(x^4-3)=0.
\label{13}
\end{equation}

The impact parameter $b$ is the radius of the shadow. The circle with the radius
$b$ in the center of the BH mass is the ``photon ring". The solution of Eq. (13) depends on $B$ which includes two independent parameters $q_m$ and $\beta$.

\section{The BH shadow}

Because QED is well proven theory, we imply that rational NLED (1) at the weak-field limit becomes QED with loop corrections (described by the Euler$-$Heisenberg Lagrangian). Therefore, the model parameter $\beta$ can be evaluated by comparing (1), at weak electromagnetic fields, with the Euler$-$Heisenberg Lagrangian \cite{Heisenberg}. Making use of series of Lagrangian (1) in the small parameter $\beta{\cal F}$ ($\beta{\cal F}\ll 1$) we obtain
\begin{equation}
{\cal L}=-{\cal F}+2\beta{\cal F}^2-6\beta^2{\cal F}^3+{\cal O}\left((\beta{\cal F})^4\right).
\label{14}
\end{equation}
The the Euler$-$Heisenberg Lagrangian (the QED Lagrangian with one loop correction) can be approximated as \cite{Gies}
\begin{equation}
{\cal L}_{EH}=-{\cal F}+c_1{\cal F}^2,~~~c_1=\frac{8\alpha^2}{45m_e^4},
\label{15}
\end{equation}
where the electron mass is $m_e=0.51~\mbox{MeV}$ and the coupling constant $\alpha=\approx 1/137$. Making the identification of Eqs. (14) and (15) up to ${\cal O}\left((\beta{\cal F})^2\right)$, we obtain
\begin{equation}
\beta=\frac{4\alpha^2}{45m_e^4}=69\times 10^{-5}~\mbox{MeV}^{-4}.
\label{16}
\end{equation}
Now, we in the position to compare the shadow size for the BH within our model with the shadow size of M87$^*$ BH detected by the EHT \cite{Event1}. We use the M87$^*$ mass $M=(6.5\pm 0.9)\times 10^9 M_\odot\approx 72.4\times10^{68}$ MeV and the magnetic mass in our model (6) with the value for the parameter $\beta$ (16). Making use of the identity $m_M=M$ and the value of the parameter $B=q_mG/\sqrt{\beta}$ in Eq. (7) and the Newton constant $G=67.9\times10^{-58}$ eV$^2$, one obtains the magnetic charge (in Gaussian units with $c=1$) and the unitless parameter $B$
\begin{equation}
q_m\approx1.6\times10^{46},~~~~B\approx 4205.
\label{17}
\end{equation}
With the help of values (17) we find the numerical solutions to Eq. (13)
\begin{equation}
x_{ph}\approx 7005,~~~~b\approx 2.5\times10^{20}~\mbox{eV}^{-1}.
\label{18}
\end{equation}
The horizon radius is the solution for the equation $f(x_h)=1-Bg(x_h)/x_h=0$ which, for $B=4205$, is $x_h\approx 4670$. Because $x_{ph}>x_h$ the BH shadow radius is defined by the photon capture radius (18).
By virtue of the value of the impact parameter (18) (the photon capture radius), one obtains the diameter of the M87$^*$ shadow in units $MG$ ($m_M=M$) within our model
\begin{equation}
\frac{2b}{m_MG}\approx 10.4.
\label{19}
\end{equation}
The diameter of the M87$^*$ shadow measured by EHT is $d_{M87^*}=11.0\pm 1.5$. Thus, within $1\sigma$ uncertainties our result (19) is in very good agreement with the experimental data for the angular size of the shadow $\delta=(42\pm3)$ $\mu$as, the distance for M87$^*$ $D=(16.8\pm 0.8)$ Mpc and the diameter of the M87$^*$ shadow (in units $MG$) $d_{M87^*}=D\delta/(MG)$. The angular size of the shadow $\delta$ is defined as the shadow diameter $2b$ divided by the distance from the BH to the observer ($D$), $\delta=2b/D$.
One can verify that the condition $d^2V/dr^2>0$ for stable orbits is satisfied for the solution obtained.

If the parameter $\beta$ is not fixed, then from Eqs. (6) and (13) we find
\[
\frac{2b}{m_MG}=\frac{8\sqrt{2}x\sqrt{x^4-3}}{\pi B\sqrt{(x^4+1)(1-Bg(x)/x)}},
\]
\begin{equation}
B=\frac{2x(x^8+6x^4-3)}{g(x)(3x^8+10x^4-9)-x^3(x^4-3)}.
\label{20}
\end{equation}
The plot of the unitless function $2b/(m_MG)$ versus $x$ is depicted in Fig. 1.
\begin{figure}[h]
\includegraphics[height=4.0in,width=4.0in]{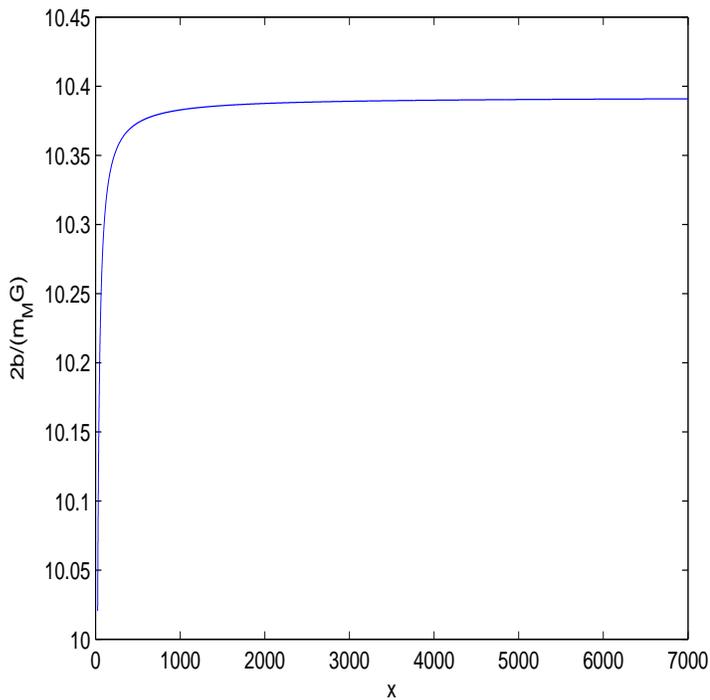}
\caption{\label{fig.1}The function $2b/(m_MG)$ versus $x$.}
\end{figure}
According to Fig. 1 the diameter of the BH shadow (in terms of $MG$) increases with increasing $x$. But $x\sim r(M\beta)^{-1/3}$ (this follows from $x=r/(\beta^{1/4}\sqrt{q_M})$ and $M=m_M$) and, therefore, the diameter of the BH shadow decreases with increasing $\beta$. The same behaviour of the diameter of the BH shadow takes place in other models of NLED coupled to GR \cite{Vagnozzi}.

\section{Conclusion}

In this paper we derived the effective metric (Eq. 9), nonlinear equations for the unitless value $x$ and the impact parameter $b$. The numerical solutions to Eq.(13), the magnetic charge and the shadow size of M87* BH were obtained. The important result is to treat M87* BH within rational NLED as a magnetically charged BH that size is in agreement with the EHT collaboration data. The next step will be the way how to distinguish this scenario from the standard one. This will be possible after receiving more precise data from the EHT collaboration.
Our result (19) for the diameter of the M87$^*$ shadow is very close to the estimation within the Schwarzschild BH, where the impact parameter (the photon capture radius) is $b_{Sch}=3\sqrt{3}(MG)$, $2b_{Sch}/(MG)\approx10.4$. The model of regular magnetic BH based on rational NLED has an advantage compared to the singular Schwarzschild BH and with BH based on Euler$-$Heisenberg electrodynamics \cite{Vagnozzi} because of the absence of singularities \cite{Krug2}. We considered the spherically symmetric spacetime because of its simplicity and ignored therefore the rotation of the BH. It worth noting that the rotational uncharged Kerr BH possesses singularities on a ring. Further we plan to find corrections to the shape of the shadow due to the spin of magnetically charged BHs based on rational NLED.


\begin{thebibliography}{999}

\bibitem{Luminet} P. Luminet, Astron. Astrophys. \textbf{75} (1979) 228-235.
\bibitem{Lu}R.-S. Lu, A. E. Broderick, F. Baron, J. D. Monnier, V. L. Fish, S. S. Doeleman et al.,
 Astrophys. J. \textbf{788} (2014) 120.
\bibitem{Gralla}S. E. Gralla, D. E. Holz and R. M. Wald, Phys. Rev. D \textbf{100} (2019) 024018.
\bibitem{Narayan}R. Narayan, M. D. Johnson and C. F. Gammie, Astrophys. J. \textbf{885} (2019) L33.
\bibitem{Falcke}H. Falcke, F. Melia and E. Agol, Astrophys. J. \textbf{528} (2000) L13.
\bibitem{Event1} Event Horizon Telescope collaboration, K. Akiyama et al., Astrophys. J. \textbf{875} (2019)
L1; ibid L2; ibid L3; ibid L4; ibid L5; ibid L6.
\bibitem{Israel}W. Israel, Phys. Rev. \textbf{164} (1967) 1776-1779.
\bibitem{Israel1}W. Israel, Commun. Math. Phys. \textbf{8} (1968) 245-260.
\bibitem{Carter}B. Carter, Phys. Rev. Lett. \textbf{26} (1971) 331-333.
\bibitem{Broderick}A. E. Broderick, T. Johannsen, A. Loeb and D. Psaltis, Astrophys. J. \textbf{784} (2014) 7.
\bibitem{Johannsen}T. Johannsen and D. Psaltis, Astrophys. J. \textbf{718} (2010) 446-454.
\bibitem{Johannsen1}T. Johannsen, A. E. Broderick, P. M. Plewa, S. Chatzopoulos, S. S. Doeleman, F. Eisenhauer
et al., Phys. Rev. Lett. \textbf{116} (2016) 031101.
\bibitem{Johannsen2}T. Johannsen, Class. Quant. Grav. \textbf{33} (2016) 113001.
\bibitem{Psaltis}D. Psaltis, Gen. Rel. Grav. \textbf{51} (2019) 137.
\bibitem{Tian}S. X. Tian, Zong-Hong Zhu, Phys. Rev. D \textbf{100}, 064011 (2019).
\bibitem{Banerjee}I. Banerjee, S. Chakraborty, S. SenGupta, Phys. Rev. D \textbf{101}, 041301 (2020).
\bibitem{Rummel}M. Rummel, C. P. Burgess, JCAP, \textbf{2020}, 051 (2020).
\bibitem{Liu}C. Liu, T. Zhu, Q. Wu, K. Jusufi, M. Jamil, M. Azreg-Aïnou, A. Wang, Phys. Rev. D \textbf{101}, 084001 (2020).
\bibitem{Wei}S.-W. Wei, Y.-C. Zou, Y.-X. Liu and R. B. Mann, JCAP \textbf{1908} (2019) 030.
\bibitem{Shaikh}R. Shaikh, Phys. Rev. D \textbf{100} (2019) 024028.
\bibitem{Davoudiasl}H. Davoudiasl and P. B. Denton, Phys. Rev. Lett. \textbf{123} (2019) 021102.
\bibitem{Bambi1}C. Bambi, K. Freese, S. Vagnozzi and L. Visinelli, Phys. Rev. D \textbf{100} (2019) 044057.
\bibitem{Safarzadeh}M. Safarzadeh, A. Loeb and M. Reid, Mon. Not. Roy. Astron. Soc. \textbf{488} (2019) L90-L93,
\bibitem{Kawashima}T. Kawashima, M. Kino and K. Akiyama, Astrophys. J. \textbf{878} (2019) 27.
\bibitem{Bar}N. Bar, K. Blum, T. Lacroix and P. Panci, JCAP \textbf{1907} (2019) 045.
\bibitem{Vagnozzi}S. Vagnozzi and L. Visinelli, Phys. Rev. D \textbf{100} (2019) 024020.
\bibitem{Dokuchaev}V. I. Dokuchaev and N. O. Nazarova, Universe \textbf{5} (2019) 183.
\bibitem{Neves}J. C. S. Neves, Eur. Phys. J. C \textbf{80}, 343 (2020).
\bibitem{Cunha}P. V.P. Cunha, C. A. R. Herdeiro, E. Radu, Universe, \textbf{5}(12), 220.
\bibitem{Banerjee1}I. Banerjee, S. Sau, S. SenGupta, Phys. Rev. D. \textbf{101}, 104057 (2020).
\bibitem{Kumar}R. Kumar, S. G. Ghosh, A. Wang, Phys. Rev.D \textbf{101}, 104001 (2020).
\bibitem{Vagnozzi1}S. Vagnozzi, C. Bambi, L. Visinelli, Class. Quant. Grav. \textbf{37}, 087001 (2020).
\bibitem{Vagnozzi2}M. Khodadi, A. Allahyari, S. Vagnozzi, D. F. Mota, arXiv:2005.05992 [gr-qc].
\bibitem{Synge} J. L. Synge, Mon. Not. Roy. Astron. Soc. \textbf{131} (1966) 463–466.
\bibitem{Penrose}R. Penrose, Phys. Rev. Lett. \textbf{14} (1965) 57-59.
\bibitem{Hawking}S. W. Hawking and R. Penrose, Proc. Roy. Soc. Lond. A \textbf{314} (1970) 529-548.
\bibitem{Penrose1}R. Penrose, Riv. Nuovo Cim. \textbf{1} (1969) 252-276.
\bibitem{Born}M. Born and L. Infeld, Proc. Roy. Soc. Lond. A \textbf{144} (1934) 425-451.
\bibitem{Heisenberg}W. Heisenberg and H. Euler, Z. Phys. \textbf{98} (1936) 714-732.
 \bibitem{Pellicer} R. Pellicer and R. J. Torrence, J. Math. Phys. \textbf{10} (1969) 1718.
\bibitem{Oliveira} H. P. de Oliveira, Class. Quant. Grav. \textbf{11} (1994) 1469.
\bibitem{Ayon1} E. Ay\'{o}n-Beato and A. Gar\'{c}ia, Phys. Rev. Lett.  \textbf{80} (1998) 5056.
\bibitem{Bronnikov1}  K. A. Bronnikov, Phys. Rev. Lett. \textbf{85} (2000) 4641.
\bibitem{Bronnikov} K. A. Bronnikov, Phys. Rev. D \textbf{63} (2001) 044005.
\bibitem{Yajima}H. Yajima and T. Tamaki, Phys. Rev. D \textbf{63} (2001) 064007.
 \bibitem{Bronnikov2} K. A. Bronnikov, G. N. Shikin, and E. N. Sibileva, Grav. Cosmol. \textbf{9} (2003) 169.
\bibitem{Burinskii} A. Burinskii and S. R. Hildebrandt, Phys. Rev. D \textbf{65} (2002) 104017.
\bibitem{Dey}T. K. Dey, Phys. Lett. B \textbf{595} (2004) 484-490.
\bibitem{Cai}R.-G. Cai, D.-W. Pang and A. Wang, Phys. Rev. D \textbf{70} (2004) 124034.
\bibitem{Allaverdizadeh}M. Allaverdizadeh, S. H. Hendi, J. P. S. Lemos and A. Sheykhi, Int. J. Mod. Phys. D \textbf{23} (2014) 1450032.
\bibitem{Olmo}G. J. Olmo, D. Rubiera-Garcia and H. Sanchis-Alepuz, Eur. Phys. J. C \textbf{74} (2014) 2804. 
\bibitem{Kruglov1}S. I. Kruglov, Phys. Rev. D \textbf{94} (2016) 044026.
\bibitem{Kruglov2}S. I. Kruglov, Ann. Phys. \textbf{378} (2017) 59-70.
\bibitem{Kruglov3}S. I. Kruglov, Ann. Phys. \textbf{383} (2017) 550-559;
\bibitem{Kruglov4}S. I. Kruglov, Ann. Phys.  \textbf{409} (2019) 167937;
\bibitem{Kruglov5}S. I. Kruglov, Ann. Phys. (Berlin) \textbf{529} (2017) 1700073;
\bibitem{Kruglov6}S. I. Kruglov, Ann. Phys. (Berlin) \textbf{528} (2016) 588-596;
 \bibitem{Kruglov7}S. I. Kruglov, Int. J. Mod. Phys. D \textbf{26} (2017) 1750075;
 \bibitem{Kruglov8}S. I. Kruglov, Int. J. Mod. Phys. A \textbf{32} (2017) 1750147;
  \bibitem{Kruglov9}S. I. Kruglov, Int. J. Mod. Phys. A \textbf{33} (2018) 1850023;
  \bibitem{Kruglov10}S. I. Kruglov, Universe \textbf{4} (2018) 66;
\bibitem{Kruglov11}S. I. Kruglov,  Eur. Phys. J. C \textbf{80} (2020) 250.
\bibitem{Krug2}S. I. Kruglov, Ann. Phys. \textbf{353} (2015) 299-306.
\bibitem{Novello} M. Novello, V. A. De Lorenci, J. M. Salim and R. Klippert, Phys. Rev. D \textbf{61} (2000) 045001.
\bibitem{Novello1} V. A. De Lorenci, R. Klippert, M. Novello and J. M. Salim, Phys. Lett. B \textbf{482} (2000) 134-140.
\bibitem{Kr} S. I. Kruglov. Dyonic and magnetic black holes with rational nonlinear electrodynamics.     \textit{Preprints} \textbf{2020}, 2020010049 (doi:10.20944/preprints202001.0049.v1).
\bibitem{Vagnozzi} A. Allahyari, M. Khodadi, S. Vagnozzi, and D. F. Mota, JCAP \textbf{2002} (2020) 003.
\bibitem{Gies} W. Ditrich and H. Gies, \textit{Probing the Quantum Vacuum} (Springer Tracts in Modern Physics, \textbf{166}, 2000).

\end{thebibliography}
\end{document}